\documentclass[twocolumn,showpacs,showkeys,prc,aps,amsmath,amssymb,10pt]{revtex4-1}
\usepackage{citesort}
\usepackage{graphicx}
\usepackage{color}
\usepackage{multirow}
\usepackage{bm}
\usepackage{mathtools}
\usepackage{subfig}
\usepackage{array}
\usepackage{float}
 
\begin{document}

\title{An effective method to accurately calculate the phase space factors for $\beta^- \beta^-$ decay}
\author{Andrei Neacsu and Mihai Horoi }
\affiliation{Department of Physics, Central Michigan University, Mount Pleasant, Michigan 48859, USA}

\begin{abstract}
\begin{center}
 {\bf Abstract}
\end{center}

 Accurate calculations of the electron phase space factors are necessary for reliable predictions of double-beta decay rates, 
 and for the analysis of the associated electron angular and energy distributions. We present an effective method to calculate
 these phase space factors that takes into account the distorted Coulomb field of the daughter nucleus, 
 yet allows one to easily calculate the phase space factors with good accuracy relative to 
 the most exact methods available in the recent literature. 
\end{abstract}
\maketitle

\section{Introduction}
Double-beta decay $(\beta\beta)$ processes are of considerable importance for the study of neutrinos.
They change the charge $Z$ of a nucleus by two units, releasing two electrons, while the mass $A$ remains unchanged.
The $\beta\beta$ decay with two associated electron antineutrinos in the final state conserves the lepton number, and is 
permitted within the Standard Model (SM). This process, called two-neutrino double-beta decay ($2\nu\beta\beta$), 
has been experimentally observed for several isotopes with transitions to both ground states and to excited states of 
the daughter nuclei \cite{Barabash2015}. Should the lepton number conservation be violated, then theories beyond the 
standard model (BSM) predict that the $\beta\beta$ decay transition could occur without antineutrinos in the final
state, called neutrinoless double-beta ($0\nu\beta\beta$), and this implies that the neutrino is a Majorana fermion \cite{SchechterValle1982}.
The $0\nu\beta\beta$ transitions have not yet been confirmed experimentally, but there are many recent experimental and 
theoretical efforts dedicated to their discovery. Recent reviews on this matter are in 
Refs. \cite{Avignone2008,Ejiri2010,Vergados2012}. There are several mechanisms that could contribute to the 
$0\nu\beta\beta$ decay rate, of which the simplest and most studied one involves the exchange of light Majorana 
neutrinos in the presence of left-handed weak interaction. Other, more complex, mechanisms include contributions from
right-handed currents \cite{Mohapatra1975,Doi1983}, and mechanisms involving super-symmetry \cite{Vergados2012,Hirsch1996}.

The phase space factors (PSF) that enter the $\beta\beta$ life-times expressions were considered for a long time as being
accurately calculated (see e.g. Refs. \cite{Doi1985,SuhonenCivitarese1998}). Recent reevaluations of the PSF, using
methods that take into account the proton distributions distorting the Coulomb field of the daughter nucleus 
\cite{Kotila2012,StoicaMirea2013,MireaPahomi2015,Stefanik2015}, have shown considerable differences in some cases when compared 
to the previous results \cite{Doi1985,SuhonenCivitarese1998}.
A very recent paper \cite{Stefanik2015} presents four of the different methods commonly used to calculate the PSF, 
and compares their results for the case of $0\nu\beta\beta$ ground state (g.s.) transitions. 
Table 1 and Fig. 2 of Ref. \cite{Stefanik2015} show that the Coulomb distortion of the electron wavefunction 
by inclusion of the finite nuclear size and electron screening effects can produce differences of up to 100\%, 
compared to the point-charge formalism of Ref. \cite{Doi1985} (see for example the $0\nu\beta\beta$ PSF $G_{08}$ of $^{150}$Nd in Ref. \cite{Stefanik2015}). 
However, taking into account the charge distributions in the daughter nuclei and 
solving numerically the Dirac equation with finite nuclear size is very slow and plagued by convergence issues. 
This makes these complex methods unattractive for the calculations of electron angular and energy distributions, such as those presented in Ref. \cite{snemo2010,HoroiNeacsu2015}.

In this paper, we propose an effective method for treating the distortion of the Coulomb field in the daughter nucleus.
This method uses the well known formalism of Ref. \cite{Doi1985}, but provides accurate results that are in good agreement with those of 
Refs. \cite{Kotila2012,StoicaMirea2013,MireaPahomi2015,Stefanik2015}. This method could be particularly useful when performing complex 
investigations involving PSF to test BSM physics due to different possible underlying mechanisms contributing to the $0\nu\beta\beta$ process.
These investigations often involve calculations of electron distributions \cite{snemo2010,HoroiNeacsu2015}, where components of the PSF enter the equations, 
and it is not possible to only use the tabulated values of Refs. \cite{Kotila2012,Stefanik2015,StoicaMirea2013,MireaPahomi2015}.

The paper is organized as follows: Section \ref{formalism} shows the formalism for $0\nu\beta\beta$ transitions to ground states, and 
for $2\nu\beta\beta$ transitions to ground and excited states. In Section \ref{method} we present our effective method for the 
treatment of the distorted Coulomb field in the daughter nucleus. Section \ref{results} is dedicated to the results, and Section \ref{conclusions} 
shows our conclusions. The Appendixes summarize the point-charge formalism from  Refs. \cite{Doi1985,SuhonenCivitarese1998}
that we adjusted  to calculate the $0\nu\beta\beta$ and 
$2\nu\beta\beta$ PSF.

\section{\label{formalism} Brief formalism of the $\beta\beta$ decay}
For the $0\nu\beta\beta$ decay, one usually writes the inverse half-life as products of electron PSF, 
nuclear matrix elements (NME) that depends on the nuclear structure of the parent and that of the daughter 
nuclei, and lepton number violation (LNV) parameters of the BSM mechanisms taken into account. 
Considering the existence of right-handed currents, one would find several additional contributions to
the decay rate \cite{Doi1983,Doi1985}. The most studied mechanism is that of the light left-handed neutrino
exchange, but other mechanisms could be of importance \cite{Vergados2012}.
One popular model that includes contributions of right-handed currents is the left-right 
symmetric model \cite{MohapatraPati1975,Senjanovic1975}. This model assumes the existence of heavy particles
that are not included in the Standard Model (SM). Within this framework, the $0\nu\beta\beta$ half-life expression is given by
\begin{eqnarray}
\label{flifetime}
\nonumber \left[ T^{0\nu}_{1/2} \right] ^{-1}  &=&  G^{0\nu}_{01} g^4_A \mid M^{0\nu}\eta_{\nu} +M^{0N}\left(\eta^L_{N_R}+\eta^R_{N_R}\right) \\
  &+& \eta_{\lambda}X_{\lambda}+\eta_{\eta}X_{\eta}\mid ^2  ,
\end{eqnarray}
where $g_A$ is the Axial-Vector coupling strength, $\eta_{\nu}$, $\eta^L_{N_R}$, $\eta^R_{N_R}$, $\eta_{\lambda}$, and $\eta_{\eta}$ are neutrino physics 
parameters defined in Ref. \cite{Barry2013}, $M^{0\nu}$  and $M^{0N}$ are the light 
and heavy neutrino-exchange nuclear matrix elements \cite{Horoi2013,Vergados2012}, and $X_{\lambda}$ 
and $X_{\eta}$ represent combinations of NME and phase space factors.
$G_{01}^{0\nu}$ is a phase space factor \cite{SuhonenCivitarese1998} that can be calculated with relatively 
good precision in most cases \cite{Kotila2012,StoicaMirea2013,Stefanik2015}. Other possible contributions, 
such as those of R-parity violating SUSY particle exchange \cite{Vergados2012,Horoi2013}, etc, are neglected here. 
With some simplifying notations the half-life expression \cite{Doi1985} (here we omit the contribution from the $\eta^L_{N_R}$ 
and $\eta^R_{N_R}$ terms, which share the same PSF as $\eta_{\nu}^2$ term, $G^{2\nu}_{01}$, and have the same energy and angular distribution as the $\eta_{\nu}$ term) is written as
\begin{flalign}
\label{lifetime}
\nonumber & \left[ T^{0\nu}_{1/2} \right]^{-1} =  g^4_A \left[ C_1 \eta^2_\nu + \textmd{cos} \phi_1 C_{2}\eta_\nu \eta_\lambda 
+ \textmd{cos}\phi_2 C_3 \eta_\nu \eta_\eta  \right. & \\
&\left. + C_4 \eta_\lambda^2 + C_5 \eta_\eta^2 + \textmd{cos}(\phi_1 - \phi_2) C_6 \eta_\lambda \eta_\eta \right] \left|M_{GT}^{0\nu}\right| ^2 ,&
\end{flalign}
where $\phi_1$ and $\phi_2$ are the relative CP-violating phases \cite{Barry2013}, and  $M_{GT}^{0\nu}$ is the Gamow-Teller part of the light left-handed neutrino-exchange NME. 
Different processes give rise to several contributions: $C_1$ comes from the left-handed leptonic and currents, 
$C_4$ from the right-handed leptonic and right-handed hadronic currents, and $C_5$ from the right-handed
 leptonic and left-handed hadronic currents. The interference between these terms is represented by the
 $C_2$, $C_3$ and $C_6$ contributions. 
Neglecting the very small tensor contributions in the mass mechanism, the
$C_{1-6}$ components are defined as products of PSF and NME \cite{Doi1985}:
\begin{subequations} \label{ci-uri}
\begin{flalign}
C_1 &= G_{01}^{0\nu} \left( 1-\chi_F \right)^2, &\\
C_2 &= \left[ G_{04}^{0\nu} \chi_{1+} + G_{03}^{0\nu} \chi_{2-} \right] \left( 1-\chi_F \right), &\\
\nonumber C_3 &=  \left[ G_{03}^{0\nu} \chi_{2+} -G_{04}^{0\nu} \chi_{1-} -G_{05}^{0\nu} \chi_P \right. &\\ 
  & \left. + G_{06}^{0\nu} \chi_R \right] \left( 1-\chi_F \right), &\\
C_4 &= \left[ G_{02}^{0\nu} \chi_{2-}^2 +\frac{1}{9}G_{04}^{0\nu}\chi_{1+}^2 -\frac{2}{9}G_{03}^{0\nu}\chi_{1+}\chi_{2-} \right], &\\
\nonumber C_5 &= G_{02}^{0\nu} \chi_{2+}^2+\frac{1}{9}G_{04}^{0\nu}\chi_{1-}^2 -\frac{2}{9}G_{03}^{0\nu}\chi_{1-}\chi_{2+} &\\ 
&+G_{08}^{0\nu}\chi_P^2 - G_{07}^{0\nu} \chi_P \chi_R + G_{09}^{0\nu} \chi_R^2 , &\\
\nonumber C_6 &= -2\left[ G_{02}^{0\nu} \chi_{2-}\chi_{2+} -\frac{1}{9}G_{03}^{0\nu}(\chi_{1+}\chi_{2+}+\chi_{2-}\chi_{1-})  \right. &\\
    &+ \left. \frac{1}{9}G_{04}^{0\nu}\chi_{1+}\chi_{1-}\right], &
\end{flalign}
\end{subequations}
with 
$$ \chi_{1\pm}=\chi_{GT_q}\pm 3\chi_{F_q} - 6\chi_{T_q} \textmd{ and } \chi_{2\pm}=\chi_{GT_\omega}\pm \chi_{F_\omega}-\frac{\chi_{1\pm}}{9}.$$

The fractions of NME are defined \cite{Doi1985} as 
$\chi_\alpha=M_\alpha/M_{GT}^{0\nu}$, with $\alpha=F,GT_\omega,F_\omega,GT_q,F_q,T_q,R$, and $P$ indicating other NME.
All these nine NME were calculated by several methods, including the interacting shell model (ISM) \cite{Retamosa1995,Caurier1996,HoroiNeacsu2015} 
and quasiparticle random phase approximation (QRPA) \cite{Muto1989}.
The light-neutrino mass mechanism $M^{0\nu}_{GT}$ and $M^{0\nu}_F$ NME have been extensively studied with many nuclear structure methods, 
such as interacting boson model (IBM-2) \cite{Barea2009,Barea2012,Barea2013,Barea2015}, interacting shell model (ISM) 
\cite{Retamosa1995,Caurier2008,MenendezPovesCaurier2009,Caurier2005,HoroiStoica2010,Horoi2013,SenkovHoroi2013,HoroiBrown2013,SenkovHoroiBrown2014,NeacsuStoica2014,SenkovHoroi2014,NeacsuHoroi2015}, 
quasiparticle random phase approximation (QRPA) \cite{Simkovic1999,Suhonen2010,Faessler2011,MustonendEngel2013,FaesslerGonzales2014}, 
projected hartree fock bogoliubov (PHFB) \cite{Rath2013}, 
energy density functional (EDF) \cite{Rodriguez2010}, and the relativistic energy density functional (REDF) \cite{Song2014} method. 
The NME calculated with different methods and by different groups still present large differences, and that has been a topic of many debates in the literature 
(see e.g. \cite{Faessler2012,Vogel2012}). Expressions for the $G_{01}^{0\nu}-G_{09}^{0\nu}$ PSF are given in \ref{app:appendix_psf0}.

For the $2 \nu \beta \beta$ process, the half-life for the transition to a state of angular momentum 
$J$ ($J=0$ or $2$) of the daughter nucleus is given to a good approximation by \cite{HoroiStoicaBrown2007}
\begin{equation}
\label{2nhl}
\left[ T^{2\nu}_{1/2} \right]^{-1} = G^{2\nu}_{(J)} g_A^4 \left| (m_ec^2)^{J+1} M^{2\nu}_{(J)} \right|^2 ,
\end{equation}
where $G^{2\nu}_{(J)}$ is a phase space factor \cite{Doi1985,SuhonenCivitarese1998,MireaPahomi2015} described in \ref{app:appendix_psf2},
$m_e$ is the electron mass, and $M^{2\nu}_{(J)}$ is the $2\nu\beta\beta$ NME, which can be calculated as \cite{SuhonenCivitarese1998,HoroiStoicaBrown2007}
\begin{equation}
 M^{2\nu}_{(J)} = \frac{1}{\sqrt{J+1}} \sum_k \frac{\left< J_f || \sigma \tau^- || 1^+_k\right> \left< 1^+_k || \sigma \tau^- || 0^+_i \right>}{\left( E_k + E_J \right)^{J+1} }  .
\label{2nnme}
\end{equation}
Here the $k$-sum is taken over the $1^+_k$ states with excitation energies $E_k$ in the intermediate odd-odd nucleus. 
$E_J = \frac{1}{2}Q_{\beta\beta}(J)+\Delta M$, where $Q_{\beta\beta}(J)$ is the Q-value for the transition to the state 
of angular momentum $J$ in the daughter nucleus, and $\Delta M$ is the difference in mass between the intermediate nucleus 
and the decaying nucleus. 

\section{\label{method}Description of the effective "screening" method}
Our approach is based of the formalism from Ref. \cite{Doi1985,SuhonenCivitarese1998} where the nuclear charge is considered point-like, 
but we replicate the effects of a finite size proton distribution distorting Coulomb field in the daughter nucleus by modifying the charge 
of the final nucleus ($Z_f$).
We multiply the $Z_f$ with a parameter, called "screening factor" ($S_f$) in what follows, to obtain an effective "screened charge" 
($Z_s=\frac{S_f}{100} Z_f$). For large enough energies, the tail of the Coulomb field plays a less significant role when compared to 
its part close to the nucleus, and the effect resembles a charge screening.
The PSF calculated with $Z_s$ for each nucleus are compared to those of Refs. \cite{Kotila2012,StoicaMirea2013,MireaPahomi2015,Stefanik2015} 
(called "data" below), which were obtained with methods that consider Dirac electron wave functions calculated with finite nuclear size and 
atomic electron screening. Refs. \cite{Kotila2012,Stefanik2015} take into account the finite nuclear size by an uniform charge distribution of radius R,
while Refs. \cite{StoicaMirea2013,MireaPahomi2015} consider a more realistic Woods-Saxon proton distribution inside the nucleus. It was shown \cite{Kotila2012} that
the atomic electron screening effect is small, of the order of 0.1\%.
The relative deviations between our results and the data, expressed in percentages 
($\Delta = 100 \left | \frac{\textmd{PSF - data}}{\textmd{data}} \right |$), 
are denoted with $\Delta G^{0\nu}_{01-09}$ for the 9 PSF of $0\nu\beta\beta$ transitions
to ground states, $\Delta{G^{2\nu}_{g.s.}}$ for PSF of $2\nu\beta\beta$ transitions to ground states, $\Delta{G^{2\nu}_{0_1^+}}$ 
for PSF of $2\nu\beta\beta$ transitions to the first excited $0^+$ state, and $\Delta{G^{2\nu}_{2_1^+}}$ 
for PSF of $2\nu\beta\beta$ transitions to the first excited $2^+$ state.

The PSF can be grouped into two classes: those that have only $s$-wave electron contributions 
$\left ( G^{0\nu}_{01}, G^{2\nu}_{g.s.},G^{2\nu}_{0_1^+},G^{2\nu}_{2_1^+}\right )$, 
and those for which $p$-wave electrons contribute $\left ( G^{0\nu}_{02} - G^{0\nu}_{09}\right )$. 
Here, we treat them separately naming them $s$-PSF and $p$-PSF, respectively. 
We consider the largest deviation ($\Delta_{max}$) between the PSF of a certain class and the corresponding data, 
and we search for the value of $S_f$ that minimizes it. 
Our goal is to maintain $\Delta_{max}\leq 10\%$. 
This value of the maximum deviation is considerably lower than the uncertainties of the NME contributing to the decay rate, 
Eq. (\ref{lifetime}).
We find that controlling the maximum deviation provides more stable and predictable results than minimizing a $\chi^2$ distribution. 

In our analysis data is selected as follows: 
$G^{0\nu}_{01-09}$ PSF are chosen from Table III of Ref. \cite{Stefanik2015}. Other recent results for $G^{0\nu}_{01}$ 
\cite{Kotila2012,StoicaMirea2013,MireaPahomi2015} are within a few percent of these values.
For $G^{0\nu}_{02}-G^{0\nu}_{09}$ there are no other results that take into account the finite size effects of the charge distribution.
The $G^{2\nu}_{g.s.}$ data is taken from Table 1 of the very recent Ref. \cite{MireaPahomi2015}, and it is in very good agreement with the results of Ref. \cite{Kotila2012}. 
For the $2\nu\beta\beta$ transitions to the first excited $0^+$ states we take the data from Table 2 of Ref. \cite{MireaPahomi2015}. 
There are four cases ($^{110}$Pd, $^{124}$Sn, $^{130}$Te, and $^{136}$Xe) of PSF in Ref. \cite{MireaPahomi2015} that are in 
significant disagreement with those of Ref. \cite{Kotila2012}. 
We do not take them into account in our analysis.
The data for $2\nu\beta\beta$ transitions to the first excited $2^+$ states is taken from Table 3 of Ref. \cite{MireaPahomi2015}.
In this case, there are three PSF values ($^{116}$Cd, $^{124}$Sn, and $^{136}$Xe) that seems to deviate significantly
from the model results. These $2\nu\beta\beta$ PSF were not confirmed by other groups, and they were often readjusted
\cite{Mirea2014up}. We do not include them in the analysis, but we compare them with our prediction in Table  \ref{tab_2p}.
The $2\nu\beta\beta$ data of $^{124}$Sn  attributed to Ref. \cite{MireaPahomi2015} in 
Tables \ref{tab_gs}-\ref{tab_2p}
was provided to us as private communications by the authors of Ref. \cite{MireaPahomi2015}.

\section{\label{results}Results and discussions}
 \begin{figure}
 \centering
 \includegraphics[width=0.95\linewidth]{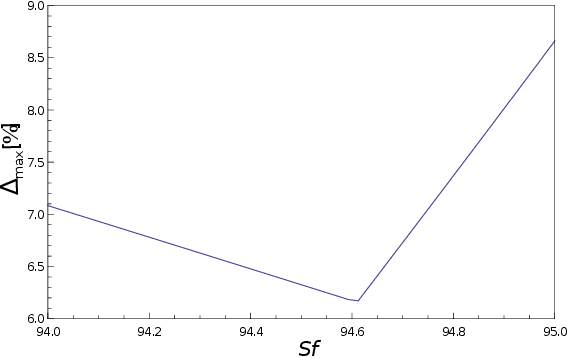}
 \caption{Fine-tuning the "screening factor" to minimize the maximum $s-$PSF deviations.}
 \label{fig_gs_conv}
 \end{figure}

\begin{table*} 
\centering
\caption{The $0\nu\beta\beta$ to ground states PSF $G^{0\nu}_{01}$ expressed in yr$^{-1}\times 10^{15}$, 
and the $2\nu\beta\beta$ to ground states $G^{2\nu}_{g.s.}$ expressed in yr$^{-1}\times 10^{20}$. 
Results are calculated with the optimal "screening factor" for $s-$PSF, $S_f=94.5$, and 
compared to those of Ref. \cite{Stefanik2015} for $0\nu\beta\beta$ and Ref. \cite{MireaPahomi2015} 
for $2\nu\beta\beta$. Second and third columns display the effective screened charge, $Z_s$, of 
the daughter nucleus and the energy of the decay, $Q_{\beta\beta}$.}
\begin{tabular}{rcc|rrc|rrc} \hline \hline	  
          &$Z_s$&$Q_{\beta\beta}$[MeV]&$G^{0\nu}_{01}$&$G^{0\nu}_{01}$\cite{Stefanik2015}&$\Delta{G^{0\nu}_{01}}$&$G^{2\nu}_{g.s.}$&$G^{2\nu}_{g.s.}$\cite{MireaPahomi2015}&$\Delta{G^{2\nu}_{g.s.}}$ \\ \hline
$^{48}$Ca &20.79&4.272 &24.55&24.83&1.1&1480.46 &1553.6  &4.7\\
$^{76}$Ge &32.13&2.039 &2.28 & 2.37&3.8&   4.51 &   4.65 &2.9\\
$^{82}$Se &34.02&2.995 &9.96 &10.18&2.1& 150.31 & 157.3  &4.4\\
$^{96}$Zr &39.69&3.35  &20.45&20.62&0.8& 642.0  & 674.4  &4.8\\
$^{100}$Mo&41.58&3.034 &15.74&15.95&1.3& 310.6  & 323.1  &3.9\\
$^{110}$Pd&45.36&2.018 &4.66 & 4.83&3.5&  12.78 &  13.25 &3.6\\
$^{116}$Cd&47.25&2.814 &16.57&16.73&1.0& 258.78 & 268.8  &3.7\\
$^{124}$Sn&49.14&2.289 &8.87 & 9.06&2.1& 51.45  &  50.4  &2.1\\
$^{130}$Te&51.03&2.527 &14.10&14.25&1.0& 142.73 & 144.2  &1.0\\
$^{136}$Xe&52.92&2.458 &14.49&14.62&0.9& 133.73 & 133.2  &0.4\\
$^{150}$Nd&58.59&3.371 &66.00&63.16&4.5&3467.53 &3539.7  &2.0\\
\hline \hline
\end{tabular} 
\label{tab_gs}
\end{table*}
\begin{table*}
\centering
\caption{PSF and their deviations for $2\nu\beta\beta$ to the first excited $0^+$ states, and $G^{2\nu}_{0_1^+}$ expressed in yr$^{-1}\times 10^{22}$.
The last two columns present PSF of Ref. \cite{Kotila2012} and their deviations. The results marked with "*" and “(*)” symbols (see text for details)
correspond to the nuclei not included in the analysis.}
\begin{tabular}{rcc|r|rr|rr}
\hline \hline	 
&$Z_s$&$Q_{\beta\beta}[MeV]$&$G^{2\nu}_{0_1^+}$&$G^{2\nu}_{0_1^+}$\cite{MireaPahomi2015}&$\Delta{G^{2\nu}_{0_1^+}}$\cite{MireaPahomi2015}&$G^{2\nu}_{0_1^+}$\cite{Kotila2012}&$\Delta{G^{2\nu}_{0_1^+}}$\cite{Kotila2012} \\ \hline
$^{48}$Ca &20.79&1.275 &    3.43& 3.52&2.6	&3.63	&5.5	\\
$^{76}$Ge &32.13&0.917 &    0.64& 0.61&5.1	&0.70	&7.7	\\
$^{82}$Se &34.02&1.508 &   41.94&41.7 &0.6	&---	&---	\\
$^{96}$Zr &39.69&2.202 & 1633.8 &1694 &3.6	&1754	&6.9	\\
$^{100}$Mo&41.58&1.904 &  562.08&570.8&1.5	&605.5	&7.2	\\
$*^{110}$Pd&45.36&0.547&  0.043&0.033&30.9	&0.048	&10.8	\\
$^{116}$Cd&47.25&1.057 &    8.00&7.59 &5.4	&8.73	&8.3	\\
$(*)^{124}$Sn&49.14&1.120&   15.09&14.1 &7.0	&---	&---	\\
$*^{124}$Sn&49.14&0.630&0.180 & --- & ---	&0.199	&9.7	\\
$*^{130}$Te&51.03&0.734&    0.69&0.55 &25.9	&0.76	&9.2	\\
$*^{136}$Xe&52.92&0.979&    3.31&2.82 &17.2	&3.62	&8.7	\\
$^{150}$Nd&58.59&2.631 &40637.5 &41160&1.3	&43290	&6.1	\\ 
\hline \hline
\end{tabular}
\label{tab_0p}
\end{table*}

\begin{table}
\centering
\caption{PSF and their deviations for $2\nu\beta\beta$ to the first excited $2^+$ states, and $G_{2^+_1}^{2\nu}$  expressed in yr$^{-1}\times 10^{21}$.
 Denoted with "*" symbol are PSF and deviations corresponding to the nuclei not included in the analysis.}
\begin{tabular}{rcc|rrc} \hline \hline	 
 &$Z_s$&$Q_{\beta\beta}[MeV]$&$G_{2^+_1}^{2\nu}$ &$G_{2^+_1}^{2\nu}$ \cite{MireaPahomi2015}&$\Delta{G_{2^+_1}^{2\nu}}$ \\ \hline
$^{48}$Ca &20.79&3.284 &   3816&4074 &6.3\\
$^{76}$Ge &32.13&1.480 &   0.40& 0.38&3.5\\
$^{82}$Se &34.02&2.219 &  71.16& 69.6&2.2\\
$^{96}$Zr &39.69&2.571 & 730.8 &742.5&2.0\\
$^{100}$Mo&41.58&2.494 & 585   & 569 &2.8\\
$^{110}$Pd&45.36&1.359 &   0.46& 0.46&0.8\\
*$^{116}$Cd&47.25&1.520&   2.11& 1.88&12.4\\
*$^{124}$Sn&49.14&1.686&   8.89& 7.63&16.5\\
$^{130}$Te&51.03&1.990 &  81.09& 79.6&1.9\\
*$^{136}$Xe&52.92&1.640&   9.03& 7.68&17.6\\
$^{150}$Nd&58.59&3.037 &31964  &30308&5.5\\ 
\hline \hline
\end{tabular}
\label{tab_2p}
\end{table}

 \begin{figure}
 \centering
 \includegraphics[width=0.95\linewidth]{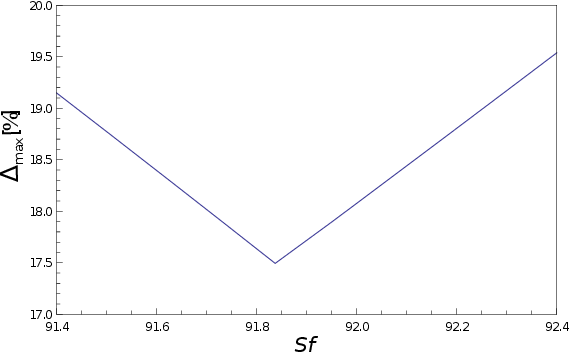}
 \caption{Same as Fig. \ref{fig_gs_conv} for $p-$PSF.}
 \label{fig_g29_conv}
 \end{figure}

 \begin{figure*}
 \centering
 \includegraphics[width=0.8\linewidth]{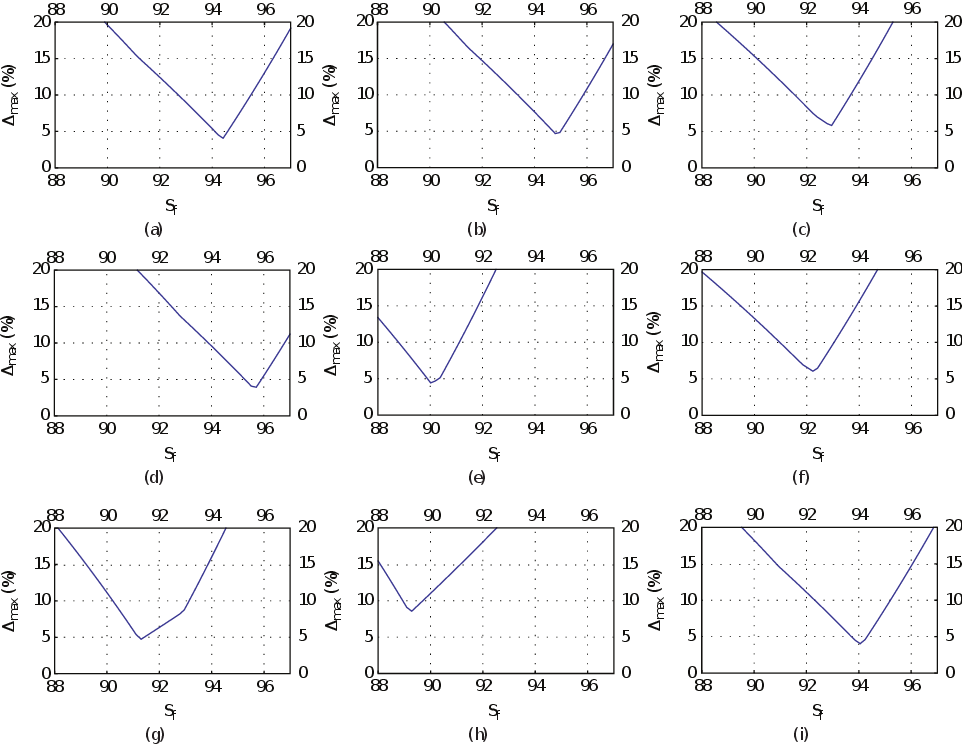}
 \caption{The behaviour of maximum deviations from the reference values of the 9 PSF involved in the $0\nu\beta\beta$ transitions to 
 ground states. Subfigures (a) to (i) correspond to $G^{0\nu}_{01}$ to $G^{0\nu}_{09}$.
 The horizontal axis represents the "screening factor" values, and the vertical axis represents the maximum deviations expressed in percentages.}
 \label{fig_g19}
 \end{figure*}
 
 \begin{table*}
\caption{The calculated $0\nu\beta\beta$ PSF ($G^{0\nu}_{01} - G^{0\nu}_{09}$) expressed in yr$^{-1}$ for the 
 decay to ground state of the 11 nuclei listed in Table \ref{tab_gs}. 
The last line shows the optimal
 "screening factor" $S_f$ for all 8 $p-$PSF ($G^{0\nu}_{02} - G^{0\nu}_{09}$). For $G_{01}$, the
 $s-$PSF optimal "screening factor", $S_f=94.5$ of Table \ref{tab_gs}, was used. Shown in the 
 last column are the maximum deviations between our calculations with the indicated parameters and the 
 results from Ref. \cite{Stefanik2015}.}
 \begin{tabular}{rlllllllllll|cc} \hline \hline
				&$^{48}$Ca&$^{76}$Ge&$^{82}$Se&$^{96}$Zr&$^{100}$Mo&$^{110}$Pd&$^{116}$Cd&$^{124}$Sn&$^{130}$Te&$^{136}$Xe&$^{150}$Nd& $S_f[\%]$& $\Delta_{max}[\%]$ \\ \hline
 $Q_{\beta\beta}$[MeV] 		&\ 4.272  &\ 2.039  &\ 2.995  &\ 3.350	&\ 3.034   &\ 2.018   &\ 2.814	 &\ 2.289   &\ 2.527   &\ 2.458	  &\ 3.371 \ &      &    \\  
 & & & & & & & & & & & & \\
$G^{0\nu}_{01}\cdot 10^{14}$	&\ 2.454  &\ 0.228  &\ 0.997  &\ 2.045	&\ 1.574   &\ 0.466   &\ 1.657	 &\ 0.887   &\ 1.410   &\ 1.449   &\ 6.600 \ & 94.5 & 4.5 \\
 & & & & & & & & & & & & \\
$G^{0\nu}_{02}\cdot 10^{14}$	&\ 16.11  &\ 0.376  &\ 3.468  &\ 8.928  &\ 5.733   &\ 0.782   &\ 5.309   &\ 1.919   &\ 3.719   &\ 3.639   &\ 30.65 \ & 95.0 & 5.0 \\
$G^{0\nu}_{03}\cdot 10^{15}$	&\ 18.45  &\ 1.233  &\ 6.671  &\ 14.50  &\ 10.72   &\ 2.548   &\ 10.96   &\ 5.254   &\ 8.853   &\ 8.967   &\ 47.87 \ & 93.0 & 6.1 \\
$G^{0\nu}_{04}\cdot 10^{15}$	&\ 5.283  &\ 0.453  &\ 2.099  &\ 4.382  &\ 3.345   &\ 0.937   &\ 3.511   &\ 1.829   &\ 2.960   &\ 3.035   &\ 14.45 \ & 95.5 & 4.2 \\
$G^{0\nu}_{05}\cdot 10^{13}$	&\ 3.134  &\ 0.559  &\ 2.011  &\ 4.139  &\ 3.464   &\ 1.337   &\ 4.003   &\ 2.427   &\ 3.694   &\ 3.895   &\ 15.27 \ & 90.0 & 4.5 \\
$G^{0\nu}_{06}\cdot 10^{12}$	&\ 3.869  &\ 0.496  &\ 1.655  &\ 2.951  &\ 2.388   &\ 87.46   &\ 2.482   &\ 1.472   &\ 2.157   &\ 2.209   &\ 7.813 \ & 92.0 & 6.6 \\
$G^{0\nu}_{07}\cdot 10^{10}$	&\ 2.790  &\ 0.268  &\ 1.161  &\ 2.432  &\ 1.885   &\ 0.566   &\ 1.984   &\ 1.052   &\ 1.663   &\ 1.703   &\ 7.799 \ & 91.0 & 6.0 \\
$G^{0\nu}_{08}\cdot 10^{11}$	&\ 1.212  &\ 0.154  &\ 0.732  &\ 1.776  &\ 1.417   &\ 0.443   &\ 1.654   &\ 0.891   &\ 1.468   &\ 1.548   &\ 7.946 \ & 89.5 & 9.3 \\
$G^{0\nu}_{09}\cdot 10^{10}$	&\ 15.97  &\ 1.172  &\ 4.647  &\ 8.471  &\ 6.399   &\ 1.863   &\ 6.131   &\ 3.211   &\ 4.884   &\ 4.878   &\ 20.09 \ & 94.0 & 4.2 \\ \hline
$p-$PSF & \multicolumn{11}{c|}{Common $S_f$ parameter for $G^{0\nu}_{02} - G^{0\nu}_{09}$} & 92.0 & 18.1 \\ \hline \hline 
\end{tabular}
 \label{tab_g0niu}
\end{table*}

For the analysis of the $s-$PSF we consider the $G^{0\nu}_{01}$ of Ref. \cite{Stefanik2015} and $G^{2\nu}_{g.s.}$, $G^{2\nu}_{0_1^+}$, 
$G^{2\nu}_{2_1^+}$ of Ref. \cite{MireaPahomi2015}. 
We find that the smallest maximum deviations from the data can be obtained using an optimal "screening factor" $S_f=94.5$. 
Fig. \ref{fig_gs_conv} shows how the maximum deviation reaches a minimum when one gets close to the optimal "screening factor".

Table \ref{tab_gs} presents $s-$PSF and their deviations $\Delta{G^{0\nu}_{01}}$ and $\Delta{G^{2\nu}_{g.s.}}$ for transitions to ground states.
The adjusted charge of the daughter nucleus is also presented, together with the $Q_{\beta\beta}$ values. 
We find very good agreement for these $s-$PSF and the data, with deviations smaller than 5\%. 
Should one consider the point-charge formalism \cite{Doi1985}, the largest deviation goes up to 40\% for the case of 
$G^{0\nu}_{01}$ of $^{150}$Nd (see, \emph{e.g.}, Table 1 columns A and D of Ref. \cite{Stefanik2015}).

Table \ref{tab_0p} shows PSF and deviations from the data for $2\nu\beta\beta$ transitions to the first excited $0^+$ states. 
The largest $\Delta{G^{2\nu}_{0_1^+}}=5.4\%$ was for $^{116}$Cd. The point-charge formalism deviations exceed 38\%.
 The PSF marked with the "*" symbol correspond to the four nuclei not included in our analysis. 
Our results for these cases can be considered as predictions for these cases that are
not yet validated by other methods.
The last two columns show the results
 of Ref. \cite{Kotila2012} and the corresponding deviations, for comparison. Ref. \cite{Kotila2012} provides no value for $^{82}$Se. 
 The case of the $^{124}$Sn is more complicated because of the 
different values used in the literature for the energy of the first excited $0^+$ state (see Ref.
\cite{HoroiNeacsu124Sn2015} for details). We include here with “*” the phase space factor corresponding 
to the $Q_{\beta\beta}$ used in Ref. \cite{Kotila2012}, and with “(*)” the phase space factor corresponding to the $Q_{\beta\beta}$ considered in Ref. \cite{Dawson2008} (see discussion in Ref. \cite{HoroiNeacsu124Sn2015}).
 The $G^{2\nu}_{2_1^+}$ PSF and their deviations are displayed in Table \ref{tab_2p}.  
 We find the largest deviation $\Delta{G_{2^+_1}^{2\nu} }=6.3\%$ for $^{48}$Ca.
 Neglecting finite nuclear size effects, one would get a deviation of 47\% for $^{150}$Nd.
Similar to the previous table, the three results excluded from the analysis are presented for comparison and marked with the "*" symbol.

When calculating the $p-$PSF, $G^{0\nu}_{02}-G^{0\nu}_{09}$ we find a different optimal "screening factor", $S_f=92$, 
corresponding to a larger maximum deviation, $\Delta_{max}=18.1\%$.
Fig. \ref{fig_g29_conv} presents the evolution of $\Delta_{max}$ close to the "optimum screening factor" for $p-$PSF. 
We attribute this larger deviations to the different kinematic factors of the nine $0\nu\beta\beta$ PSF (see Eqs. (\ref{b-factors})).

To further minimize the deviations, we obtain eight optimal "screening factors" corresponding to $G^{0\nu}_{02}-G^{0\nu}_{09}$, 
as seen in Fig. \ref{fig_g19}. The best results are presented in Table \ref{tab_g0niu}, where we show the optimal "screening factor", 
the maximum deviations, and the PSF values. 
The last line presents the optimal "screening factor" that minimizes the deviations of all the $p-$PSF.
Alongside $G^{0\nu}_{02}-G^{0\nu}_{09}$, we display the $G^{0\nu}_{01}$ obtained with the optimal "screening factor" common for all $s-$PSF.

\section{\label{conclusions}Conclusions}
In this paper we present an effective method to calculate the phase space factors of the $\beta^-\beta^-$ transitions, 
which can provide results close to those of methods that consider the finite size of the proton charge  and the atomic electron screening. It modifies the point-charge formalism of Refs. \cite{Doi1985,SuhonenCivitarese1998}, by considering a constant multiplicative ”screening factor” for the charge of the daughter nucleus.
The main advantage of our method consists in its simplicity given its accuracy, and its potential to
be extended to calculations of the energy and angular electron distributions needed for the analysis of the contributions of the right-handed currents to the $0\nu\beta\beta$ decay.

Our method works well for PSF of $0\nu\beta\beta$ and $2\nu\beta\beta$ transitions to ground stated, and also for $2\nu\beta\beta$ transitions to the first excited $0^+$ and $2^+$ states. 
For PSF where only $s-$wave electrons contribute, an effective "screening factor", $S_f=94.5$, was obtained.
Using this $S_f$ value one finds a maximum deviation of 6.3\% between our results and other results in recent literature
\cite{Stefanik2015,MireaPahomi2015,StoicaMirea2013,Kotila2012}. In the case of the PSF where $p-$wave electrons contribute, 
we obtained another optimal "screening factor", $S_f=92$. This corresponds to a maximum deviation of 18.1\% between our results and those of Ref. \cite{Stefanik2015}. 
We attribute this large deviation to the kinematic factors of Eqs. (\ref{b-factors}).
The deviations are greatly reduced, to less than 10\%, when considering individual "screening factors" for each specific PSF ($G^{0\nu}_{02}-G^{0\nu}_{09}$). 
It is remarkable that in the case of the $G_{08}$, the original point-charge formalism \cite{Doi1985} PSF deviates by over 100\% for $^{150}$Nd, while it is significantly reduced in our model.
Similar spectacular reductions are found for other $p-PSF$. 
We also provide predictions for the PSF of some isotopes, which can be also used as guidance in cases of disagreement between the more precise methods.

In addition, using $S_f=92$ one gets the largest maximum deviation of 18.1 for all neutrinoless double-beta decay PSF, $G^{0\nu}_{01}-G^{0\nu}_{09}$. This information is relevant for the calculation of the two-electron energy and angular distributions \cite{HoroiNeacsu2015}.

We conclude that this method is well suited for fast and accurate calculations of the $\beta\beta$ decay PSF, with uncertainties much lower
than those of the associated NME. 
One could envision to further reduce these PSF uncertainties by considering a mass-dependent “screening factor”. A Mathematica notebook that can be used to obtain all
these phase space factors can be downloaded from \cite{NeacsuHoroiMNbook}.

\begin{acknowledgments}
The authors acknowledge useful discussions with S. Stoica and M. Mirea.
Support from  the NUCLEI SciDAC Collaboration under
U.S. Department of Energy Grant No. DE-SC0008529 is acknowledged.
MH also acknowledges U.S. NSF Grant No. PHY-1404442
\end{acknowledgments}

\appendix
\section{\label{app:appendix_psf0}$0\nu\beta\beta$ PSF expressions}
The $0\nu\beta\beta$ PSF are calculated by integrating over the energy of one electron $(\epsilon_1)$ using the following expression adopted from Eq. (A.27) of Ref. \cite{SuhonenCivitarese1998}:
\begin{flalign}
 \label{psfs-form}
&G_{0k}=\frac{a_{0\nu}}{\textmd{ln}2(m_e R)^2}\int\limits_1^{T+1} b_k F_0(Z,\epsilon_1)F_0(Z,\epsilon_2)\omega_{0\nu}(\epsilon_1) \text{d}\epsilon_1, &
\end{flalign}
where $R$ is the nuclear radius ($R=r_0 A^{1/3}$, with $r_0=1.2$ fm), 
$\epsilon_2=T+2-\epsilon_1$, $p_{1,2}=\sqrt{\epsilon_{1,2}^2-1}$, $T=\frac{Q_{\beta\beta}}{m_e}$, and
$\omega_{0\nu}(\epsilon_1)= p_1 p_2 \epsilon_1 \epsilon_2$.
The constant  $a_{0\nu}$ is
\begin{equation} \label{a0}
a_{0\nu}=\frac{\left( G_F \textmd{cos}\theta_c \right) ^4 m_e^9}{32\pi^5} = 1.94 \times 10^{-22}\ \textmd{yr}^{-1}.
\end{equation}
We use $G_F=1.1663787\times 10^{-5} GeV^{-2}$ for the Fermi constant, and $\textmd{cos}\theta_c=0.9749$ for the Cabbibo angle.
The Fermi function used in Eq. (\ref{psfs-form}) is given by
\begin{eqnarray}\label{psffunction}
\nonumber F_{0}(Z_s,\epsilon)&=&\left[ \frac{2}{\Gamma(2\gamma_1+1)}\right] ^2 \times  (2pR)^{2(\gamma_1-1)}\\
                             &\times& |\Gamma(\gamma_1+iy)|^2 e^{\pi y},
\end{eqnarray}
where
\begin{equation}
\gamma_1=\sqrt{1-(\alpha Z_s)^2}, \qquad y=\alpha Z_s \epsilon /p.
\end{equation}
Here $\alpha$ is fine structure constant, and $Z_s$ represents the "screened" charge of the final nucleus.
The kinematic factors $b_k$ are defined as:
\allowdisplaybreaks[1]
\begin{subequations}\label{b-factors}
\begin{eqnarray}
b_1&=1,   \\
b_2&=\frac{1}{2}\left( \frac{\epsilon_1 \epsilon_2 -1}{\epsilon_1 \epsilon_2}\right) (\epsilon_1-\epsilon_2)^2,  \\
b_3&= (\epsilon_1-\epsilon_2)^2 / \epsilon_1 \epsilon_2 ,  \\
b_4&= \frac{2}{9} \left( \frac{\epsilon_1 \epsilon_2 -1}{\epsilon_1 \epsilon_2}\right) ,  \\
b_5&= \frac{4}{3} \left(\frac{(T+2)\xi}{2r_A\epsilon_1\epsilon_2} -\frac{\epsilon_1 \epsilon_2 +1}{\epsilon_1 \epsilon_2}\right), \\
b_6&= \frac{4(T+2)}{r_A\epsilon_1\epsilon_2} ,\\
b_7&= \frac{16}{3} \frac{1}{r_A\epsilon_1\epsilon_2}\left( \frac{\epsilon_1\epsilon_2+1}{2r_A} \xi-T-2 \right) ,\\
\nonumber b_8&= \frac{2}{9} \frac{1}{r_A^2\epsilon_1\epsilon_2}\left[(\epsilon_1\epsilon_2+1)(\xi^2+4r_A^2) \right.  \\
 &- \left. 4r_A \xi(T+2) \right] , \\
b_9&= \frac{8}{r_A^2}\left( \frac{\epsilon_1\epsilon_2+1}{\epsilon_1\epsilon_2} \right) ,
\end{eqnarray}
\end{subequations}
with $\xi=3\alpha Z_s+r_A(T+2)$ and $r_A=m_e R$.

\section{\label{app:appendix_psf2}$2\nu\beta\beta$ PSF expressions}
Using the formalism from Ref. \cite{SuhonenCivitarese1998}, we write the $2\nu\beta\beta$ PSF for a final state of angular momentum
$J$ ($J=0,2$) as integrals over the energies of the two emitted electrons
\begin{equation}
 G^{2\nu}_{(J)}=g_J \int_1^{T+1} F_0(Z_s,\epsilon_1)p_1\epsilon_1 I_J(T,\epsilon_1)\text{d}\epsilon_1, 
\end{equation}
with $I_J$ 
\begin{eqnarray}
\nonumber I_J(T,\epsilon_1)&=&\int_1^{T+2-\epsilon_1}F_0(Z_s,\epsilon_2)p_2\epsilon_2 \\ 
&\cdot& f_J(T+2-\epsilon_1-\epsilon_2)^{2+J}\text{d}\epsilon_2.
\end{eqnarray}
Here, $\epsilon_2=T+2-\epsilon_1$, $p_{1,2}=\sqrt{\epsilon_{1,2}^2-1}$, $T=\frac{Q_{\beta\beta}}{m_e}$, and $F_0$ is defined in Eq. (\ref{psffunction}).
In the case of $J=0$, we have $f_0=1$ and $g_0=3.78\times10^{-25}\text{\ yr}^{-1}$. For $J=2$, then $f_2=(\epsilon_1-\epsilon_2)^2$ and $g_2=g_0/7$.
\bibliography{dbd}
\end{document}